\documentclass[12pt]{article}
\usepackage{amsfonts,psfig,epsfig,latexsym,amscd,multicol,theorem,}
\newcommand{\R}{{\mathbb{R}}} 
\newcommand{\Z}{{\mathbb{Z}}} 
 
\newcommand{\C}{{\mathbb{C}}} 
\newcommand{\cx}{{{\mathbb{C}}^\times}}

\newcommand{\CP}{{\mathbb{C}}{{P}}} 
\newcommand{\RP}{{\mathbb{R}{{P}}}} 

\newcommand{\beq}{\begin{equation}} 
\newcommand{\eeq}{\end{equation}} 
\newcommand{\bea}{\begin{eqnarray}} 
\newcommand{\eea}{\end{eqnarray}} 
\newcommand{\ra}{\rightarrow} 
 
\newcommand{\hra}{\hookrightarrow} 
 
\newcommand{\cd}{\partial} 
\newcommand{\wt}{\widetilde} 
 
\newcommand{\M}{{\sf M}} 

\newcommand{\nid}{\noindent} 
\newcommand{\orot}{{\cal O}} 
 
\newcommand{\rat}{{\sf Rat}} 
\newcommand{\req}{{\sf Rat}^{eq}}

\newcommand{\ol}{\overline} 
\newcommand{\xv}{{\bf x}}

\theoremstyle{plain} 
\newtheorem{thm}{Theorem} 
 
\newtheorem{prop}[thm]{Proposition} 
 
\newtheorem{conj}[thm]{Conjecture} {\theorembodyfont{\rmfamily}

} 
\newcommand{\news}{\setcounter{equation}{0}} 
\newcommand{\comment}[1]{}

\begin{document} 

\title{Slow equivariant lump dynamics on the two sphere} 
\author{J.A. McGlade\thanks{E-mail: {\tt jmcglade@maths.leeds.ac.uk}}\,\,  and 
J.M. Speight\thanks{E-mail: {\tt speight@maths.leeds.ac.uk}} \\
School of Mathematics, University of Leeds\\
Leeds LS2 9JT, England} 

\date{} 

\maketitle 

\begin{abstract}
The low-energy, rotationally equivariant dynamics of $n$ $\CP^1$ lumps on $S^2$ 
is studied within the approximation of geodesic motion in the moduli space of 
static solutions $\req_n$. The volume and curvature properties of $\req_n$ are 
computed. By lifting the geodesic flow to the completion of  an $n$-fold cover 
of $\req_n$, a good understanding of nearly singular lump dynamics within this 
approximation is obtained.
\end{abstract}

\maketitle

\section{Introduction}\label{sec:intro} 
\news
The $\CP^1$ model in $2+1$
dimensions is a field theory of Bogomol'nyi type, analogous in many
respects to the Yang-Mills-Higgs and abelian Higgs models.  It has a
topological lower bound on energy, saturated by solutions of a first order
self-duality equation. These solutions may be interpreted as topological
solitons, called lumps, analogous to monopoles and vortices. They have
various physical interpretations in theoretical high energy and condensed
matter physics. If space is a Riemann surface $\Sigma$, then static lumps
are holomorphic maps $\Sigma\ra\CP^1$, the Cauchy-Riemann condition
playing the role of the self-duality equation. The most fruitful approach
to understanding the dynamics of $n$ moving lumps is, following Ward
\cite{war}, to restrict the field dynamics to $\M_n$, the moduli space of
degree $n$ static lumps. This is the geodesic approximation originally
proposed by Manton for monopole dynamics \cite{man}. It works well for
vortex and monopole dynamics \cite{gibman,sam,stu1,stu2}, though it lacks
a rigorous underpinning for lumps. As is well known, the reduced dynamics
amounts to geodesic motion in $(\M_n,\gamma)$ where $\gamma$ is the $L^2$
metric, defined by the restriction to $T\M_n$ of the kinetic energy
functional of the field theory. One important difference between lumps and
monopoles or vortices is that $(\M_n,\gamma)$ is geodesically incomplete
in the lump case \cite{sadspe}, so the approximation predicts that lumps
may collapse and form singularities in finite time.

In reducing to the
geodesic approximation, we replace a nonlinear hyperbolic PDE (the field
equation) by a finite system of nonlinear ODEs (the geodesic equation in
$\M_n$).This is clearly a much simpler system in principle. It is still
highly nontrivial to study its solutions, however, principally because it
is usually impossible to obtain explicit formulae for the metric $\gamma$.
The same is true for monopoles and vortices. For these systems,
interesting progress has been made by imposing extra rotational symmetries
on the geodesic problem, so as to reduce it to low-dimensional
submanifolds of $\M_n$
\cite{housut}. In the present paper, we apply this technique to $\CP^1$
lumps moving on $\Sigma=S^2$, concentrating particularly on the behaviour
of geodesics close to the singularities where lumps collapse. The $\CP^1$
model is more usually formulated on domain $\Sigma=\C$. This is a bad
choice from our viewpoint since the $L^2$ metric is undefined due to the
presence of non-normalizable zero modes \cite{war} (though one can study
geodesic motion on the leaves of a foliation of $\M_n$ on which these bad
zero modes are frozen \cite{lee}).  This problem is absent when $\Sigma$
is a compact Riemann surface. The choice $\Sigma=S^2$ is particularly
natural because then $\M_n$ (though not $\gamma$) coincides with the
$\Sigma=\C$ moduli space. Noting that $\CP^1\cong S^2$, if we choose
stereographic coordinates $z,W$ on domain and codomain respectively, then
a degree $n$ holomorphic map is simply 
\beq 
\label{*} 
\phi:z\mapsto
W(z)=\frac{a_0+ a_1z+\cdots+a_nz^n}{b_0+b_1z+\cdots+b_nz^n}=
\frac{p(z)}{q(z)} 
\eeq
where $p(z)$ and $q(z)$ have no common roots and at
least one of $a_n,b_n$ is nonzero. So $\M_n=\rat_n$, the space of degree
$n$ rational maps \cite{wood}. There is a natural open inclusion
$\rat_n\hra\CP^{2n+1}$, namely 
\beq
W(z)\mapsto [a_0,a_1,\ldots,a_n,b_0,b_1,\ldots,b_n]
\eeq
whence $\rat_n$ inherits the
structure of a complex manifold. $\rat_n$ is noncompact since it omits
from $\CP^{2n+1}$ the complex codimension 1 variety on which $p$ and $q$
share roots. As $\phi$ approaches this missing set, one or more lumps
collapse to infinitely thin spikes and disappear. It is known that
$\gamma$ is K\"ahler with respect to this complex structure \cite{spe}.  
See \cite{bap,spe} for a comprehensive survey of the geometric properties
of $(\rat_1,\gamma)$.

In the next section we identify in each $\rat_n$ a
totally geodesic submanifold $\req_n$, topologically cylindrical,
consisting of those $n$-lumps invariant under a certain $SO(2)$ action.  
We compute the induced metric on $\req_n$, also denoted $\gamma$, and the
total volume of $(\req_n, \gamma)$, which turns out to be finite and,
somewhat surprisingly, independent of $n$. In
section \ref{lifmet} we study the lift of $\gamma$ to the obvious $n$-fold
cover of $\req_n$, itself cylindrical. We show that the lifted metric
extends to a metric on $S^2$ which is $C^0$ if $n\geq 2$, $C^1$ if $n\geq
3$ and $C^2$ if $n\geq 4$, and deduce the total Gauss curvature of
$\req_n$ for $n\geq 1$. There is strong numerical evidence that
$(\req_n,\gamma)$ may be isometrically embedded as a surface of revolution
in $\R^3$, and we construct this surface numerically for small $n$.
Finally, in section \ref{geoflo} we study the geodesic problem on
$(\req_n,\gamma)$ by lifting it to the $n$-fold cover. This allows us, in
particular, to gain a good understanding of near singular
geodesics.

\section{The geometry of $\req_n$} \label{georat}
\news
There is a natural isometric action of $G=SO(3)\times SO(3)$ on 
$(\rat_n,\gamma)$
descending from the usual action of $SO(3)$ on $S^2\subset\R^3$, namely
\beq
(\orot_1,\orot_2):\phi\mapsto \orot_1\circ \phi\circ\orot_2^{-1}
\eeq
where we have used $\orot_i$ to denote both an element of $SO(3)$ and its
action on $S^2$ \cite{spe}. Given any subgroup (indeed, subset) $K$ of
$G$, the fixed point set $\rat_n^K$ of $K$ in $\rat_n$ is, if a
submanifold, a totally geodesic submanifold of $(\rat_n,\gamma)$:
geodesics which start on and tangential to $\rat_n^K$ remain on $\rat_n^K$
for all subsequent time \cite{rom}. Consider the following subgroup
$K\cong SO(2)$: 
\beq K=\{(R(n\alpha),R(\alpha)):\alpha\in\R\},\quad\mbox{where}\quad R(\alpha)  
 =\left(\begin{array}{ccc} \cos\alpha&-\sin\alpha&0\\
\sin\alpha&\cos\alpha&0\\ 0&0&1 \end{array}\right).
\eeq
Let us denote its fixed point set $\req_n$. For later convenience, we also 
define a $SO(2)$ subgroup of purely spatial rotations: 
\beq \label{k0} K_0=\{(0,R(\alpha)):\alpha\in\R\}.
\eeq
In terms of stereographic coordinates, the action of $K$ is 
\beq
\label{A} W(z)\mapsto e^{in\alpha}W(e^{-i\alpha}z).
\eeq
We may split $\rat_n$ into $U_0$, the
subset on which $b_0\neq 0$ and its complement. On $U_0$, we may uniquely
write $W(z)$ in the form 
\beq
W(z)=\frac{a_0+a_1z+\cdots+a_nz^n}{1+b_1z+\cdots+b_nz^n}.
\eeq
If $W(z)\in
U_0\cap\req_n$ then for all $z$ and
$\alpha$
\bea
\frac{a_0e^{in\alpha}+a_1e^{(n-1)i\alpha}z+\cdots+a_nz^n}{1+b_1z
e^{-i\alpha}+\cdots+b_ne^{-in\alpha}z^n}
&=&\frac{a_0+a_1z+\cdots+a_nz^n}{1+b_1z+\cdots+b_nz^n}
\eea
by (\ref{A}),
and hence 
\beq
a_0=a_1=\cdots=a_{n-1}=b_1=b_2=\cdots=b_n=0,\quad a_n\neq 0.
\eeq
Any rational map in the complement of $U_0$ may be uniquely written
\beq
W(z)=\frac{1+a_1z+\cdots+a_nz^n}{b_1z+\cdots+b_nz^n},
\eeq
since $a_0,b_0$ cannot both vanish, by the no common roots condition.
Hence if $W(z)\in \req_n$ and $W(z)\notin U_0$, 
then for all $z$ and $\alpha$ 
\beq 
\frac{e^{in\alpha}+a_1e^{(n-1)i\alpha}z+\cdots+a_nz^n}{b_1z
e^{-i\alpha}+\cdots+b_ne^{-in\alpha}z^n}=\frac{1+a_1z+\cdots
+a_nz^n}{b_1z+\cdots+b_nz^n}
\eeq
which has no solution. Hence
$\req_n\subset U_0$:
\beq
\req_n=\{az^n:a\in\cx\}.
\eeq
Clearly $\req_n$ is a
noncompact complex submanifold of $\rat_n$ of complex dimension 1,
biholomorphic to $S^2\backslash\{0,\infty\}$.

\begin{figure}[htb]
\centering
\begin{tabular}{ccc}
\includegraphics[scale=0.3333]{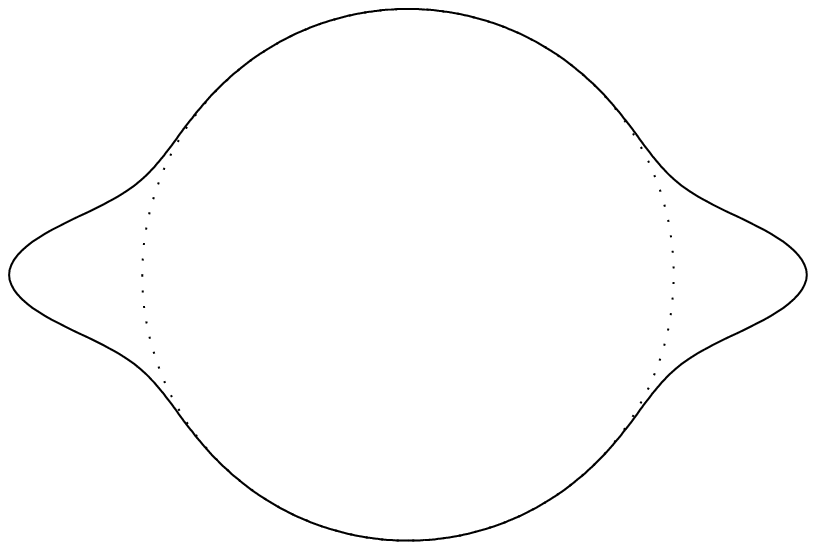}
&
\includegraphics[scale=0.3333]{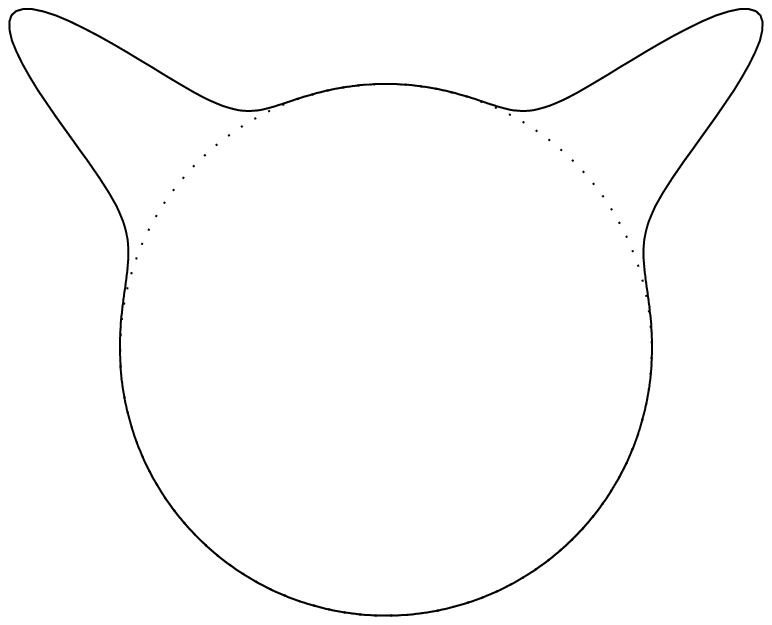}  
&
\includegraphics[scale=0.3333]{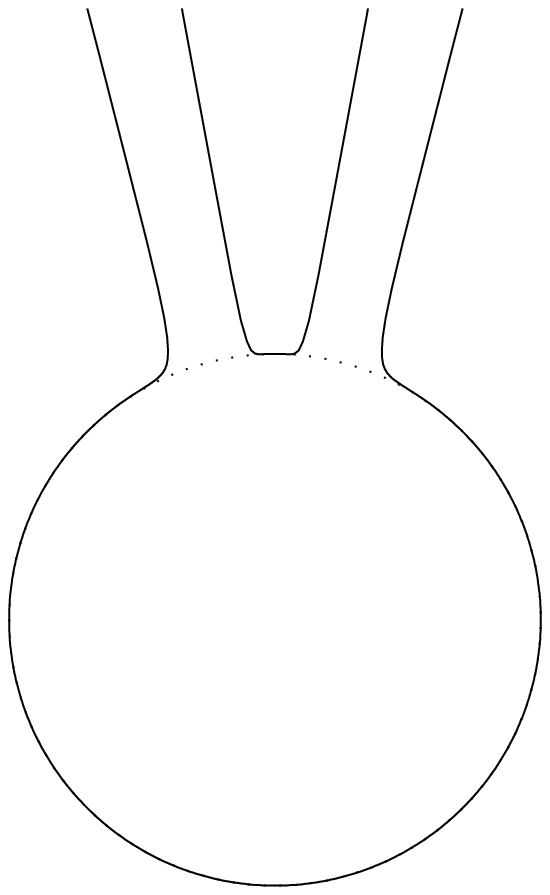} 
\end{tabular}
\caption{The energy density  ${\cal E}$ of 
$W(z)=\chi z^5$ with $\chi=1$, $1/50$ and $1/50000$ respectively. Depicted
are vertical cross sections through the graph of ${\cal E}$, plotted 
radially outwards as a non-negative function on $S^2$. The 
complete graphs are
rotationally symmetric about the vertical axis.}
\label{fig1}
\end{figure}

Physically, $\req_n$ should
be thought of as the space of coincident $n$-lumps located at either the
north or the south pole of the domain $S^2$. If $a=\chi e^{i\psi}$, then
$\chi\in\R^+$ describes the shape of the $n$-lump, while $\psi$ is its
internal phase.  The case $n=1$ was described in \cite{spe1}, so let us
assume $n\geq 2$.  The energy density ${\cal E}=\frac{1}{2}|d\phi|^2$
is $K_0$ invariant, ${\cal E}(z)={\cal E}(e^{-i\alpha}z)$, hence
independent of $\psi$, and is localized in a band centred on a circle
of constant latitude, as illustrated in
figure \ref{fig1}. Note that if $\chi\gg 1$ ($\chi\ll 
1$), the energy
accumulates towards the South pole (North pole), although ${\cal E}$ vanishes
identically at the poles themselves. One should bear in mind
that geodesics in $\req_n$ correspond to $n$-lump motions in which the
shape varies in this one-parameter family and the internal phase
simultaneously varies. The coincident lump position occupies only the two
polar values, though the band of maximum energy density does move up and
down smoothly.

Note that $K$ invariance is an admissible equivariance
constraint for the full field equation also. If we let $z=re^{i\varphi}$,
then the $\CP^1$ field equation is 
\beq
\frac{4}{(1+r^2)^2}\left[W_{tt}-\frac{2\bar{W}W_t^2}{1+|W|^2}\right]=
W_{rr}+\frac{W_r}{r}+\frac{W_{\varphi\varphi}}{r^2}-
\frac{2\bar{W}}{1+|W|^2}\left(W_r^2+\frac{W_\varphi^2}{r^2}\right)
\eeq
which supports solutions within the $K$ invariant
ansatz
\beq
W(r,\theta,t)=r^na(r,t)e^{in\varphi}
\eeq
for any $n\in\Z$. While
the complex valued function $a(r,t)$ is $C^1$, nonvanishing and has limits
at $r=0,\infty$ such solutions have degree $n$. We may regard geodesic
flow in $(\req_n,\gamma)$ as the geodesic approximation to this symmetry
reduced field dynamics, or as a symmetry reduction of the geodesic
approximation to the unreduced field dynamics.

The metric on $\req_n$ is $K_0$ invariant and hermitian, so 
\beq 
\label{jsg1}
\gamma=F(\chi)(d\chi^2+\chi^2d\psi^2) 
\eeq
for some smooth positive
function $F$. Let $\sigma:S^2\ra S^2$ denote the isometry $z\mapsto
z^{-1}$ (rotation by $\pi$ about the $x_1$ axis), and $\hat{\sigma}$
denote the corresponding isometry of $\rat_n$, that is, $\phi\mapsto
\sigma\circ\phi\circ\sigma^{-1}$. Since $\hat{\sigma}$ preserves $\req_n$,
in coordinates $\hat{\sigma}:(\chi,\psi)\mapsto (\chi^{-1},-\psi)$, it is
an isometry of $\req_n$, so from equation
 (\ref{jsg1}),
\beq
\label{isom}
\hat{\sigma}^*\gamma=\chi^{-4}F(\chi^{-1})(d\chi^2 
+\chi^2d\psi^2)=\gamma\quad \Rightarrow\quad F(\chi^{-1})\equiv \chi^4 
F(\chi).
\eeq
It suffices, therefore, to understand the geometry of the ``hemisphere'' 
of $\req_n$ where $0<\chi\leq 1$. To deduce $F(\chi)$, we must compute 
the squared $L^2$ norm of the zero mode $\cd/\cd\chi \in 
T_{(\chi,0)}\req_n$, that is, twice the initial kinetic energy of the 
field $W(z ,t)=(\chi+t)z^n$:
\beq
\label{B}
F(\chi)=\int_\C\frac{dzd\ol{z}}{(1+|z|^2)^2}\frac{|\dot{W}(z,0)|^2}{(1 + 
|W(z,0)|^2)^2}=2\pi\int_0^\infty\frac{dr}{(1 
+ r^2)^2}\frac{r^{2n+1}}{(1+\chi^2 r^{2n})^2}.
\eeq
To be consistent with previous work, we have given both domain and 
codomain the metric $(1+|z|^2)^{-1}dzd\ol{z}$, or equivalently, radius 
$\frac{1}{2}$. The $L^2$ metric for maps between spheres of radii $R_1$ 
and $R_2$ is easily deduced from this:
\beq
\gamma'=16R_1^2R_2^2\gamma.
\eeq
The even function $F:\R\ra \R^+$ defined in (\ref{B}) is smooth by, for 
example, repeated application of Lemma 2.2 from \cite{spe}. Since the 
integrand in (\ref{B}) is rational, $F(\chi)$ can be computed explicitly, 
in principle, for any $n\in\Z^+$, though in practice the expressions 
become so complicated as to be useless as $n$ increases. The integral 
formula (\ref{B}) turns out to be far more useful than the explicit 
expressions in any case. A striking illustration of this 
is

\begin{prop}\label{prop1}
$\req_n$ has volume $\pi^2$, independent of $n$.
\end{prop}

\nid {\it Proof:}\bea {\rm Vol}(\req_n)&=&\int_0^{2\pi}d\psi\int_0^\infty 
d\chi\, \chi F(\chi)= 4\pi^2\int_0^\infty 
d\chi\int_0^\infty\frac{dr}{(1+r^2)^2}\frac{\chi r^{2n+1}}{ (1+\chi^2 
r^{2n})^2}\nonumber\\
&=&4\pi^2\int_0^\infty\frac{dr\,  r}{(1+r^2)^2}\int_0^\infty d\chi\,  r^n\, 
\frac{r^n\chi}{(1+(r^n\chi)^2)^2}\nonumber \\ 
\label{C} &=&4\pi^2\left[\int_0^\infty\frac{d\alpha\, \alpha}{(1+ 
\alpha^2)^2}\right]^2= \pi^2,\nonumber
\eea
where we have applied Tonelli's Theorem \cite{priest}.$\quad\Box$\\

Of more direct consequence for the geodesic flow on $\req_n$ is an
understanding of the singularity of $\gamma$ as $\chi\ra 0$, hence, by the
isometry $\hat{\sigma}$, also as $\chi\ra \infty$. Such understanding is
obtained by lifting $\gamma$ to the $n$-fold cover of
$\req_n$.

\section{The lifted metric} \label{lifmet}

\news

There is a natural $n$-fold cover of $\req_n\cong\cx$ by $\cx$ itself, 
namely $\pi:c\mapsto
c^n$. In terms of polar coordinates $c=\rho e^{i\lambda}$,
$\pi:(\rho,\lambda)\mapsto (\chi,\psi)=(\rho^n,n\lambda)$. The lifted
metric $\wt{\gamma}=\pi^*\gamma$ on $\cx$ is 
\beq
\wt{\gamma}=\wt{F}(\rho)(d\rho^2+\rho^2d\lambda^2)\quad\mbox{where} \quad
\wt{F}(\rho)=n^2\rho^{2n-2}F(\rho^n).
\eeq
In fact, rather than deduce an integral formula for $\wt{F}$ from that for 
$F$, it is easier to compute $\wt{F}(\rho)$ directly as the squared 
$L^2$ norm of the zero mode $\cd/\cd\rho$ in the family $W(z)=(\rho z)^n$, 
\beq
\label{D} \wt{F}(\rho)=\pi
n^2\int_0^\infty\frac{ds}{(\rho^2+s)^2}\frac{s^n}{(1+s^n)^2}
\eeq
where we
have used the substitution $s=(\rho|z|)^2$. Note that $\pi(1/c)=1/\pi(c)$,
so $\hat{\sigma}:(\rho,\lambda)\mapsto (\rho^{-1},-\lambda)$ is an
isometry of $\wt{\gamma}$, and
hence
\beq
\label{jsg2}
\wt{F}(\rho^{-1})\equiv\rho^4\wt{F}(\rho)
\eeq
just as for $F(\chi)$. The integrand in (\ref{D}) is globally bounded on
$(0,\infty)$, independent of $\rho$, by $s^n(1+s^n)^{-2}$, which is
Lebesgue integrable if $n\geq 2$. Hence, by the Lebesgue dominated
convergence theorem (LDCT \cite{ldct})
\beq
\lim_{\rho\ra 0}\wt{F}(\rho)=\pi
n^2\int_0^\infty ds\lim_{\rho\ra\infty}
\frac{1}{(\rho^2+s)^2}\frac{s^n}{(1+s^n)^2} =\pi
n^2\int_0^\infty\frac{ds\, s^{n-2}}{(1+s^n)^2}
\eeq
which is finite and positive for $n\geq 2$. It follows that $\wt{\gamma}$ 
extends to a $C^0$ metric $\ol{\gamma}$ on $S^2=\cx\cup\{0,\infty\}$, 
smooth away from $0$ and $\infty$. We suspect that $\ol{\gamma}$ is never 
(i.e.\ is for no $n\in\Z^+$) a smooth metric on $S^2$, but is $C^k$ provided $n\geq k+2$.  
For our purposes it will suffice to prove this for $k=1$ and
$k=2$.

\begin{prop}\label{prop3}The $C^0$ metric
$\ol{\gamma}=\wt{F}(\rho)(d\rho^2 +\rho^2d\lambda^2)$ on $S^2$
is $C^1$ if $n\geq
3$ and $C^2$ if $n\geq 4$.\end{prop}

\noindent {\it Proof:} Since
$\ol{\gamma}$ is smooth away from $\{0,\infty\}$ and $\hat{\sigma}$ is an
isometry it suffices to check that $\ol{\gamma}$ is $C^k$, $k=1,2$, at
$\rho=0$. So $\ol{\gamma}$ is $C^1$ if $\lim_{\rho\ra 0}\wt{F}'(\rho)=0$,
and is $C^2$ if, in addition, $\lim_{\rho\ra
0}\wt{F}''(\rho)-\wt{F}'(\rho)/\rho=0$. Now
\beq\wt{F}'(\rho)=-4\pi 
n^2\rho\int_0^\infty\frac{ds}{(\rho^2+s)^3}\frac{s^n}{(1+ s^n)^2}=:-4\pi
n^2\rho f(\rho).
\eeq
The integrand of $f$ is dominated by
$s^{n-3}(1+s^n)^{-2}$ which is integrable if $n\geq 3$. Hence
$\lim_{\rho\ra 0}f(\rho)=\int_0^\infty ds\, s^{n-3}(1+s^n)^{-2}<\infty$ by
the LDCT, so $\lim_{\rho\ra 0}\wt{F}'(\rho)=0$ as required.  Further, 
\beq
\wt{F}''(\rho)-\frac{1}{\rho}\wt{F}'(\rho)=-4\pi n^2\rho f'(\rho)
\eeq
and
\beq \lim_{\rho\ra 0}f'(\rho)=\lim_{\rho\ra
0}-6\rho\int_0^\infty\frac{ds}{ (\rho^2+s)^4}\frac{s^n}{(1+s^n)^2}=0
\eeq
if $n\geq 4$ by appeal, once again, to the LDCT.\hfill$\Box$
\vspace{0.25cm}

This $C^2$ lift property has immediate consequences
for the curvature properties of $\req_n$. Let $\kappa$ and $\wt{\kappa}$
be the Gauss curvatures of $(\req_n,\gamma)$ and $(\cx,\wt{\gamma})$.  
Since $\pi$ is by definition a local isometry,
$\wt{\kappa}=\kappa\circ\pi$. If $n\geq 4$ then $\wt{\gamma}$ extends to a
$C^2$ metric on $S^2$, compact, so $\wt{\kappa}$, and hence $\kappa$, must
be bounded in this case. This should be contrasted with $(\req_1,\gamma)$
whose Gauss curvature is unbounded above. We may also compute the total
Gauss curvature of $\req_n$ exactly:

\begin{prop}\label{prop4}
The total Gauss curvature of $(\req_n,\gamma)$ is, for $n\geq 1$, $$ 
\int_{\req_n}\kappa=\frac{4\pi}{n}. $$
\end{prop}

\noindent {\it Proof:} Let $\Delta\subset\cx$ be the wedge $ 
\Delta=\{\rho e^{i\lambda}\, :\, \rho\in\R^+,\, 
0\leq\lambda<\frac{2\pi}{n}\}. $ Note that the local isometry 
$\pi:\cx\ra\req_n$ maps $\Delta$ bijectively onto $\req_n$. 
Hence
\beq
\int_{\req_n}\kappa=\int_\Delta\wt{\kappa}=\frac{1}{n} 
\int_\cx\wt{\kappa}
\eeq
by $SO(2)$ invariance of $\wt{\gamma}$. If $n\geq 4$, the total Gauss 
curvature of $(S^2,\ol{\gamma})$ is $4\pi$ since $\ol{\gamma}$ is 
sufficiently regular to apply the Gauss-Bonnet theorem. The total Gauss 
curvature of $(\cx,\gamma)$ is also $4\pi$ since $S^2\backslash\cx$ has 
measure $0$, and the result follows. To cover the cases $n=1,2,3$, one 
must resort to direct computation. Since
\beq 
\wt{\kappa}=-\frac{1}{\rho\wt{F}(\rho)}\frac{d\,\, }{d\rho}\left( 
\frac{\rho\wt{F}'(\rho)}{2\wt{F}(\rho)}\right)
\eeq
we have that
\beq 
\int_\cx\wt{\kappa}=4\pi\int_0^1d\rho\, \rho\wt{F}(\rho)\, 
\wt{\kappa}(\rho) 
=-4\pi\left[\frac{\rho\wt{F}'(\rho)}{2\wt{F}(\rho)}\right]_0^1
\eeq
where we have used the isometry $\hat{\sigma}$ to reduce the $\rho$ 
integral to $(0,1]$. Differentiating the identity (\ref{jsg2}) at 
$\rho=1$ shows that $\wt{F}'(1)=-2\wt{F}(1)$, whence the result follows 
provided
\beq
\lim_{\rho\ra 0}\frac{\rho\wt{F}'(\rho)}{2\wt{F}(\rho)}=0.
\eeq
We have already noted 
that $\lim_{\rho\ra 0}\wt{F}(\rho)$ exists and is nonzero for $n\geq 2$, 
so it remains to show that $\lim_{\rho\ra 0}\rho\wt{F}'(\rho)=0$. This 
follows from the proof of Proposition \ref{prop3} for $n\geq 3$, and may 
be checked easily for $n=2$ by computing $\rho\wt{F}'(\rho)$ explicitly 
(using, for example, Maple) and evaluating the limit by hand. The case 
$n=1$ again requires us to calculate $\rho\wt{F}'(\rho)$ explicitly, but 
now also $\wt{F}(\rho)$, take the ratio and then take the limit (using, 
for example, Maple again). \hfill $\Box$
\vspace{0.25cm}

The qualitative behaviour 
of geodesic flow on a surface depends crucially on the sign of $\kappa$. 
In this connexion we make \begin{conj}\label{conj5}For all $n\geq 1$ 
$(\req_n,\gamma)$ has positive Gauss curvature, and may be isometrically 
embedded as a surface of revolution in $\R^3$.\end{conj}\noindent There 
is strong numerical evidence for Conjecture \ref{conj5}. Assume that such 
an embedding $\xv:\req_n\ra\R^3$ does exist
\beq
\xv(\chi,\psi)=(\alpha(\chi),\beta(\chi)\cos\psi,\beta(\chi)\sin\psi).
\eeq
We may construct its generating curve by equating $\gamma$ with the 
induced metric on 
$\xv(\req_n)\subset\R^3$,
\beq
(\alpha'(\chi)^2+\beta'(\chi)^2)d\chi^2+\beta(\chi)^2d\psi^2=F(\chi) 
(d\chi^2+\chi^2d\psi^2).
\eeq
This fixes $\beta(\chi)=\chi\sqrt{F(\chi)}$. To construct $\alpha(\chi)$ 
we solve the 
ODE
\beq
\label{E}
\frac{d\alpha}{d\chi}=\sqrt{F(\chi)}\sqrt{1-\left(1+\frac{\chi 
F'(\chi)}{2 F(\chi)}\right)^2}
\eeq
with initial data $\alpha(1)=0$. Clearly, the solution exists 
whilever
\beq
\label{F}-1\leq 1+\frac{\chi F'(\chi)}{2 F(\chi)}\leq 1,
\eeq
which we find numerically holds true for all $\chi$ for 
$n=1,2,\ldots,6$. Inequality (\ref{F}) has a nice geometric 
interpretation: let $\xi(\chi)$ be the angle between the $x_1$ axis and 
the tangent to the generating curve at $(\alpha(\chi),\beta(\chi))$. Then 
$\sin\xi(\chi)$ is precisely the function bounded in (\ref{F}), so the 
generating curve exists precisely where $-1\leq\sin\xi(\chi)\leq 1$.

\begin{figure}[htb]
\centering
\includegraphics[scale=0.9]{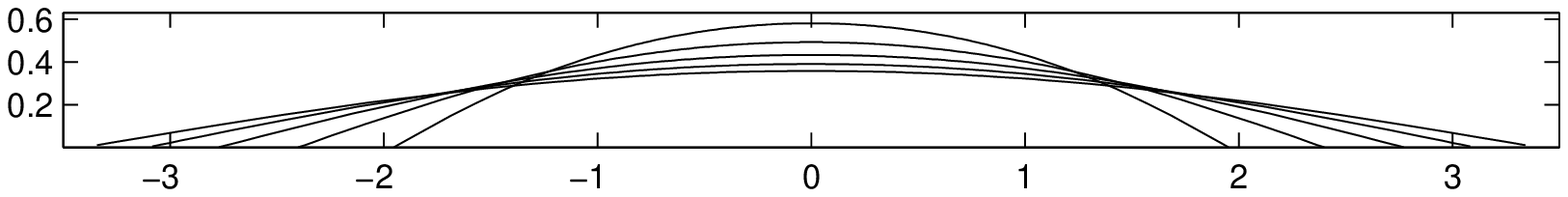}
\caption{Generating curves for $\req_n$, $n=2,\ldots 6$}
\label{fig2}
\end{figure}

We 
have solved (\ref{E}) numerically for $n=2,\ldots,6$, the resulting 
generating curves being diplayed in figure \ref{fig2}. Note that each 
curve is concave
down indicating that the surface it generates has positive Gauss 
curvature. If one changes parameters $\chi\mapsto\rho=\chi^\frac{1}{n}$, 
one finds 
\beq
\sin\xi(\chi)=\frac{\rho\wt{F}'(\rho)}{2n\wt{F}(\rho)}+\frac{1}{n} \,\, 
\stackrel{\chi\ra 0}{\longrightarrow}\,\, \frac{1}{n}\qquad (n\geq 1)
\eeq
by the argument used to prove Proposition \ref{prop4}. Hence $\req_n$, 
$n\geq 2$, has conical singularities of deficit 
angle
\beq
2\pi(1-\sin\xi(0))=2\pi(1-\frac{1}{n})
\eeq
at $\chi=0$ and 
$\chi=\infty$. This gives an alternative interpretation of the proof of 
Proposition \ref{prop4} in terms of the local Gauss-Bonnet theorem 
applied to the embedded surface of revolution \cite{localgb}:
\beq
\int_{\req_n}\kappa=2\pi(\sin\xi(0)-\sin\xi(\infty))=\frac{4\pi}{n}.
\eeq

\section{The geodesic flow} \label{geoflo}
\news

Consider the one parameter family of geodesics in $(\req_n,\gamma)$ with 
initial data $a(0)=1$, $\dot{a}(0)=e^{i\alpha}$, 
$\alpha\in[0,\frac{\pi}{2}]$. This family contains all geodesics, up to 
isometries and time rescaling. A convenient way to construct such a 
geodesic is to lift the initial data to the covering space, $c(0)=1$, 
$\dot{c}(0)=e^{i\alpha/n}$, solve the geodesic equation in 
$(\cx,\wt{\gamma})$, then project, $a(t)=(\pi\circ c)(t)=c(t)^n$. Since 
$\pi$ is a local isometry, $a(t)$ is the required geodesic. The advantage 
of this is that, for $n\geq 4$, $\wt{\gamma}$ extends to a $C^2$ metric 
$\ol{\gamma}$ on $S^2$, which is just regular enough to ensure that 
the geodesic in $(S^2,\ol{\gamma})$ exists for all time (by compactness) 
and depends continuously on the initial data. The point is that the 
geodesic equations involve only first derivatives of the metric 
coefficients, so if these coefficients are $C^2$, the flow function for 
the geodesic equation is $C^1$, hence locally Lipschitz, which is the 
minimal requirement for local existence, uniqueness and continuous 
dependence of solutions of an ODE system. So the lifting procedure allows 
one to construct reliably geodesics in $(\req_n,\gamma)$ which approach 
arbitrarily close to the singularities at $\chi=0,\infty$, and even to 
define an unambiguous continuation of the singular geodesic ($\alpha=0$) 
(which travels along the curve $\psi=0$ from $\chi=0$ to $\chi=\infty$ in 
finite time by the estimate of \cite{sadspe}) beyond both the future and 
past singularities. In the lifted picture, the ``singular'' points 
$\rho=0$ and $\rho=\infty$ are not special, and the geodesic family 
varies continuously as it approaches and hits them.

Let the closest 
approach of $|c(t)|$ to $0$ for the $\alpha$ geodesic be $\delta >0$, 
very small. This is easily computed as a function of $\alpha$ using 
angular momentum and energy conservation. For $\delta$ sufficiently 
small, $c(t)$, being $C^2$, will be well approximated by a straight line 
on the $2\delta$ disk centred on $0$. Hence the projected geodesic 
$a(t)=c(t)^n$ will wind around the singularity $\chi=0$ $(n-1)/2$ times 
before exiting the $(2\delta)^n$ disk. To describe the corresponding 
field dynamics $W(z,t)$, we shall think of a configuration as a smooth 
distribution of classical spins over physical space $S^2$, as in the 
Heisenberg model of a ferromagnet. While $a(t)$ is in the $(2\delta)^n$ 
disk, the spins are all aligned almost exactly downwards except in a 
small neighbourhood of the north pole, where they vary rapidly (in space) 
in a charge $n$ bubble. Their energy is thus highly concentrated towards 
the north pole. As $c(t)$ traverses the $2\delta$ disk, the spins precess 
rapidly $(n-1)/2$ times about the north-south axis. The configuration 
then spreads out before reforming at the south pole and undergoing a 
similar rapid precession, and so on, indefinitely. In the limit 
$\delta\ra 0$, one obtains an extended geodesic in which no spin 
precession occurs, but the configuration pinches to a point singularity 
at one pole, then spreads out to pinch at the opposite pole. There is a 
discontinuous phase flip (rotation of each spin by $\pm\pi$ about the 
north-south axis) associated with each pinch if $n$ is odd, but not if 
$n$ is even.

\begin{figure}[htb]
\centering
\begin{tabular}{ccc}
\includegraphics[scale=0.38]{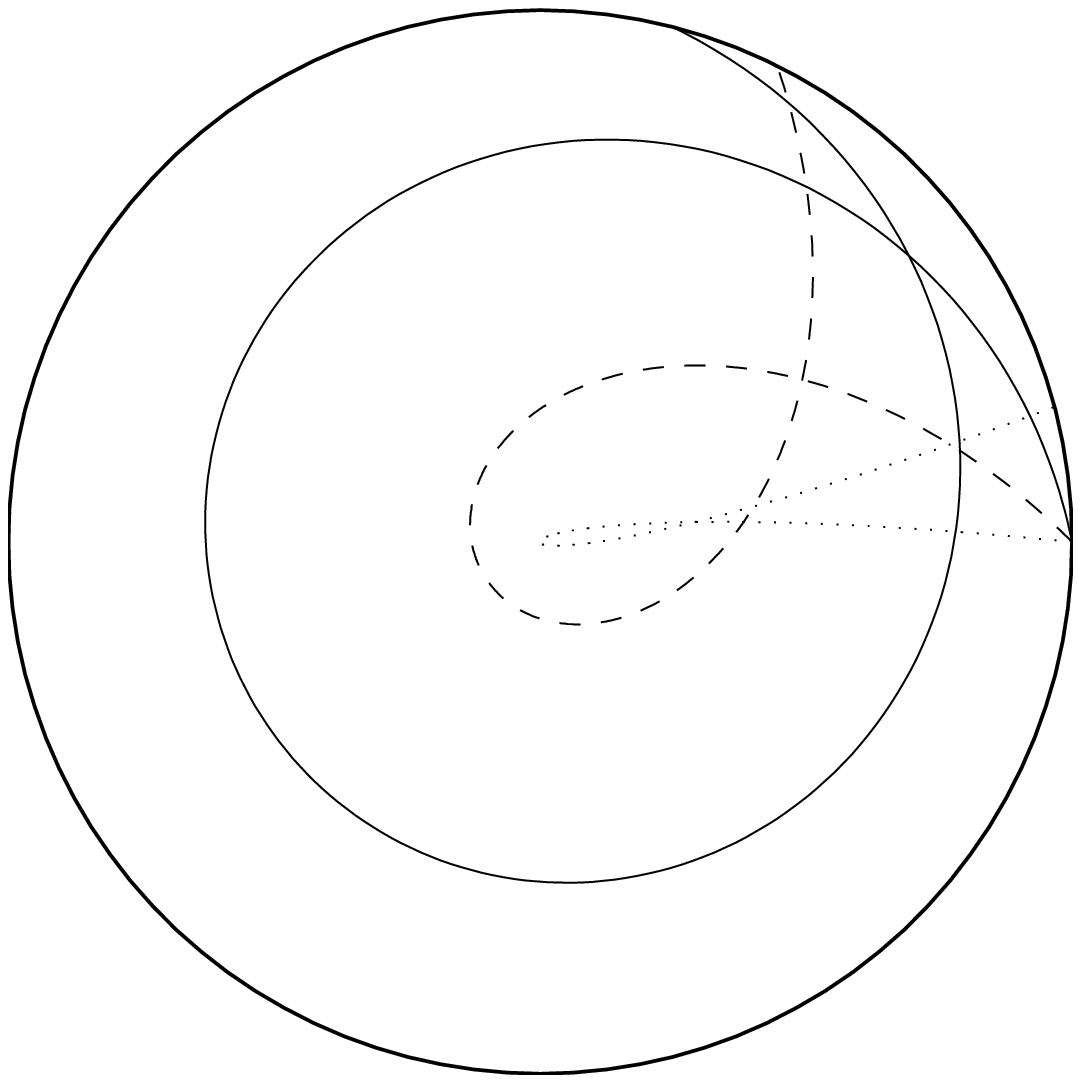}
&
\includegraphics[scale=0.38]{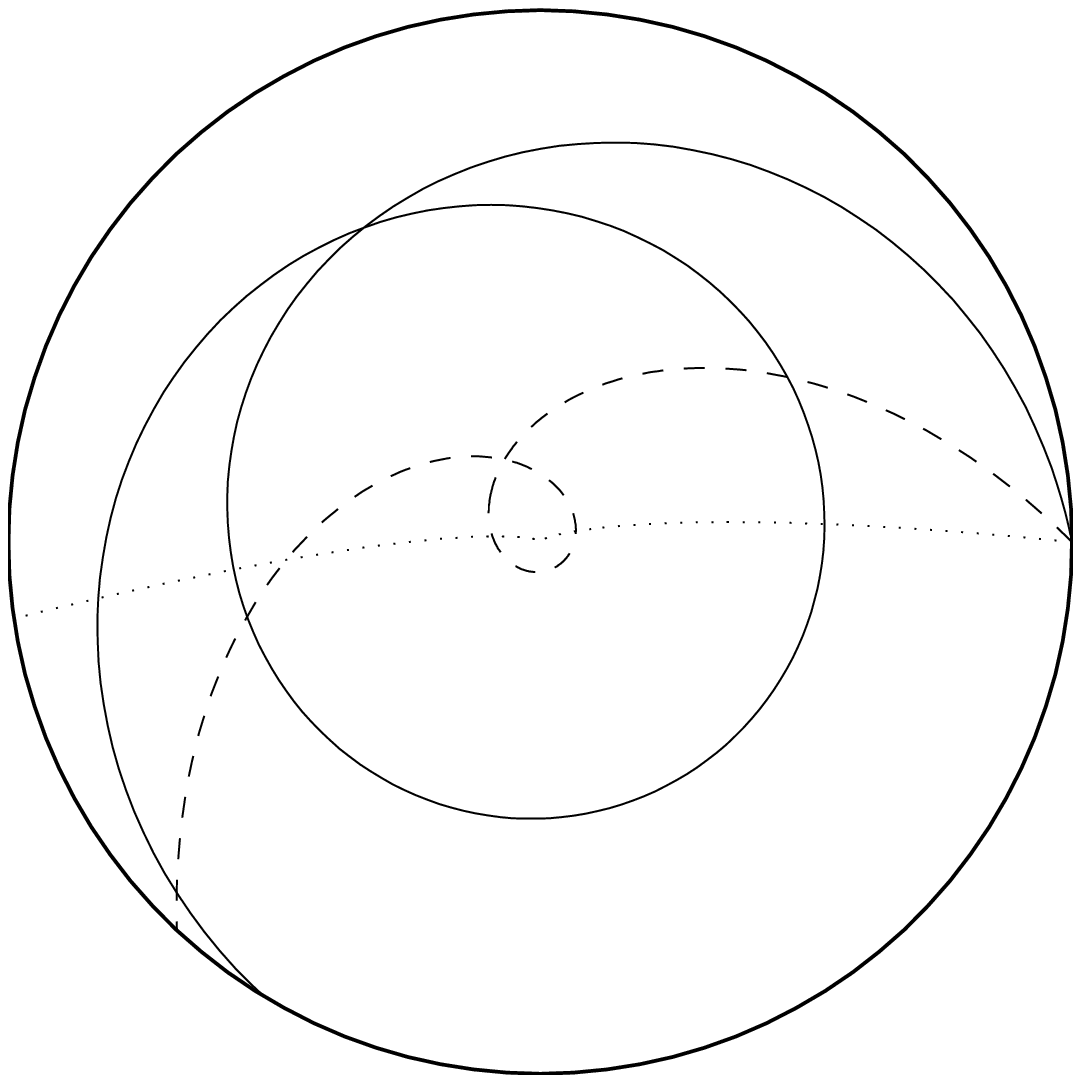}
&
\includegraphics[scale=0.38]{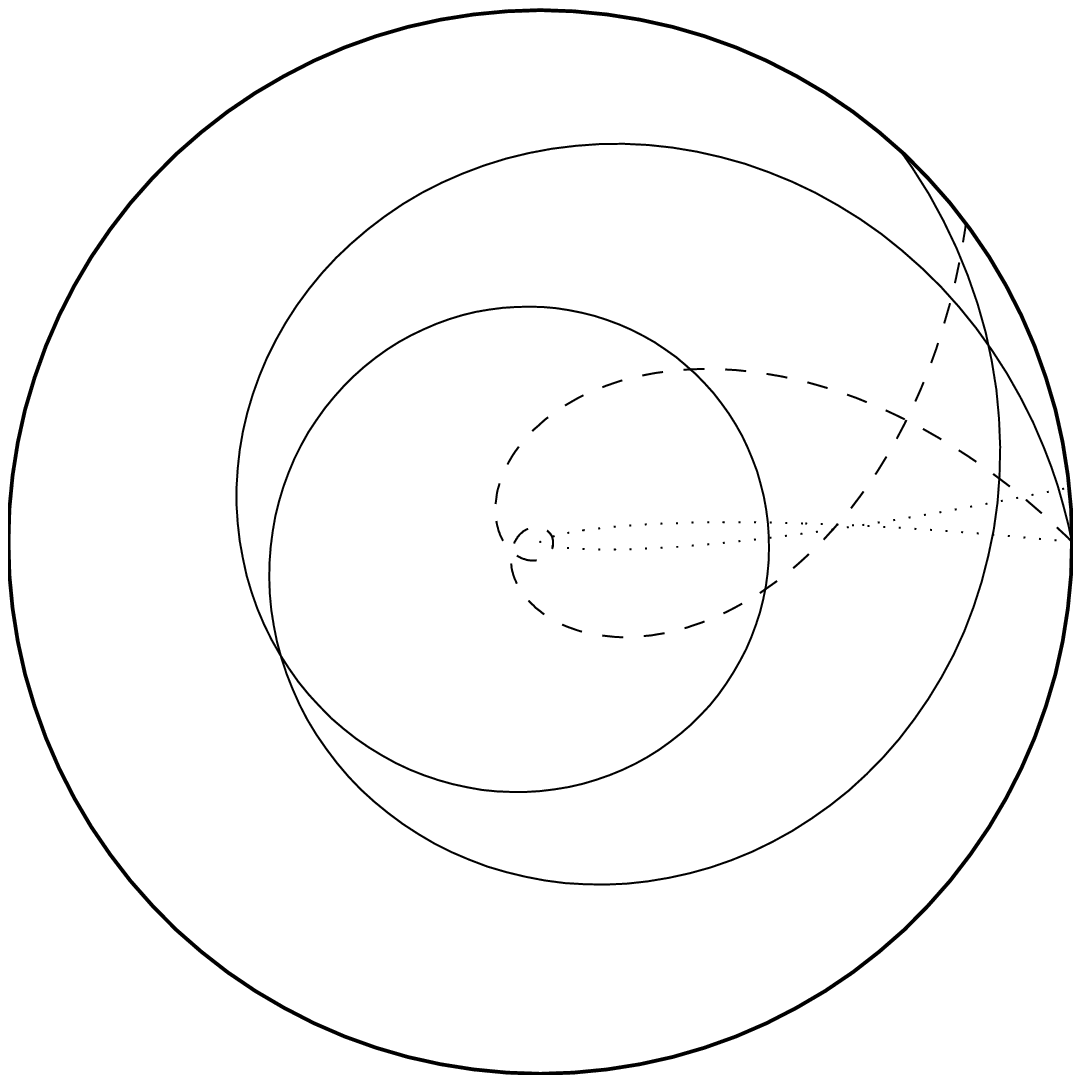}\\

\includegraphics[scale=0.38]{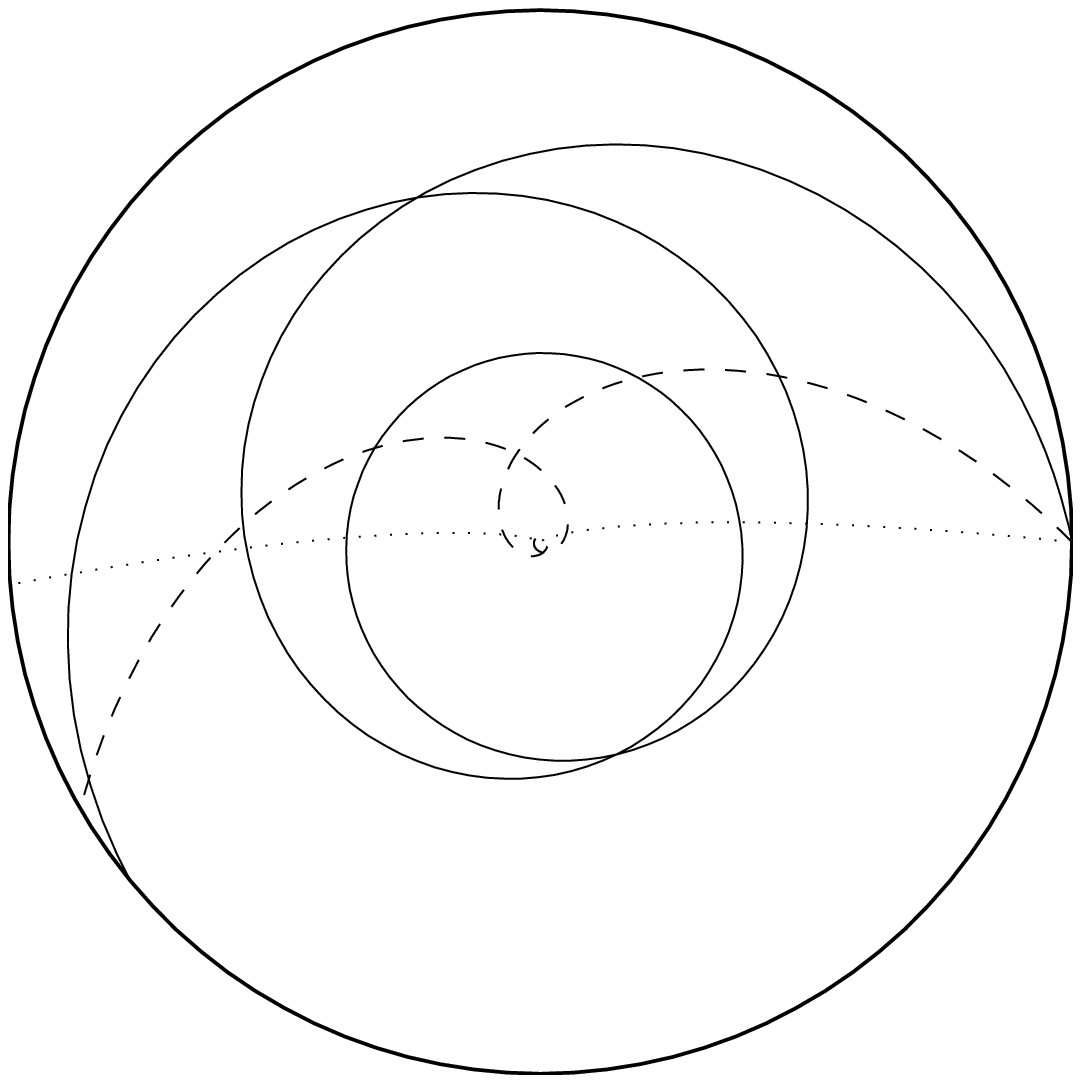}
&
\includegraphics[scale=0.38]{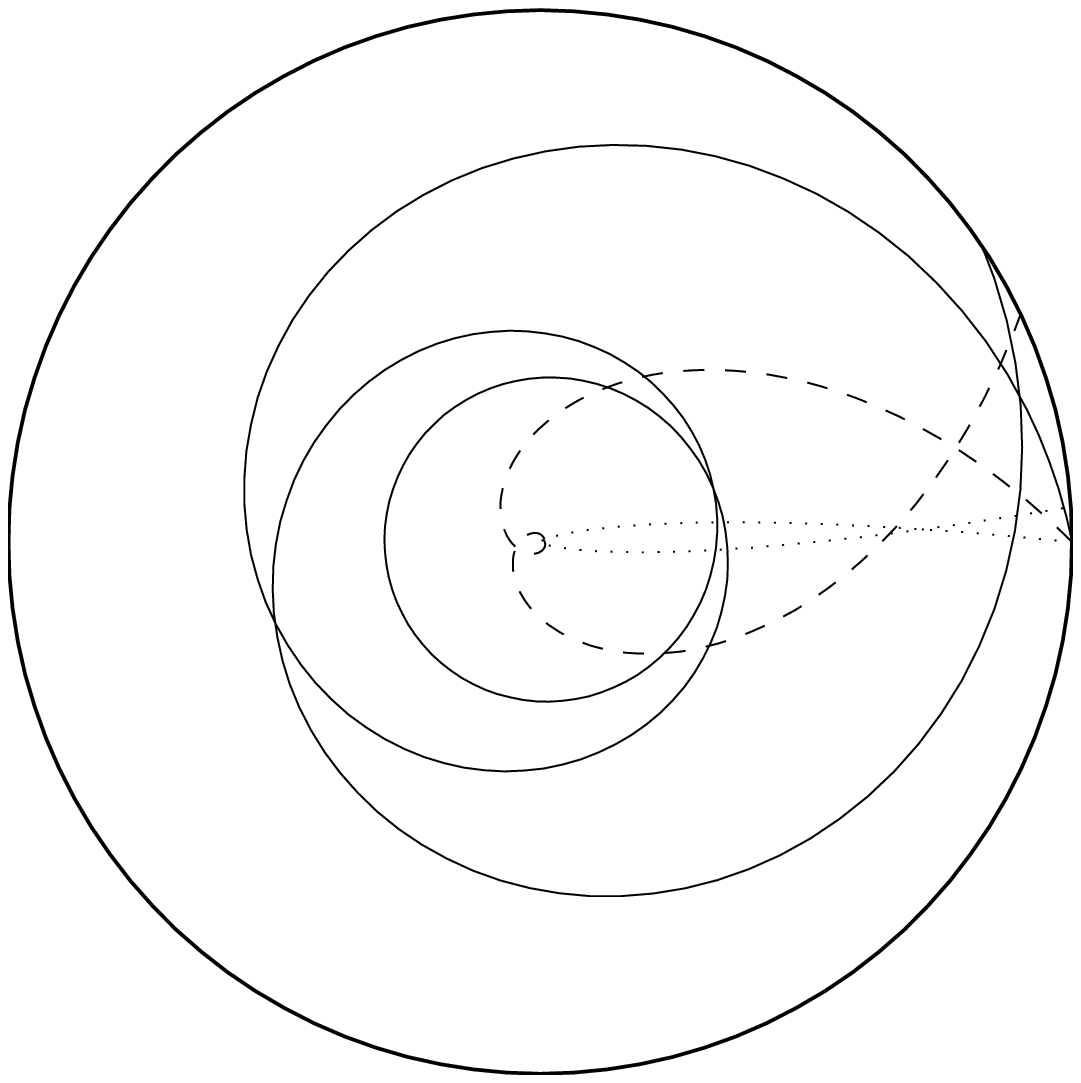}
&
\end{tabular}
\caption{Geodesics in the disk $\chi\leq 1$ in $\req_n$
when $n=2,3,4,5,6$}
\label{fig3}
\end{figure}

The above description is confirmed by numerical solution of 
the lifted geodesic problem. The equations were solved using a 4th order 
Runge Kutta method with variable time step. Energy and angular momentum 
were conserved to within $10^{-5}\,\%$. Figure \ref{fig3} shows 
the projected geodesics in various cases. Although we can only prove 
global existence and continuous dependence of all lifted geodesics for 
$n\geq 4$, the lifting procedure seems to work well also for $n=2$ and 
$n=3$. This is not surprising given the presence of conical singularities 
in these cases, as explained in section \ref{lifmet}. Of course, it is 
questionable whether geodesics which approach the singularities extremely 
closely really do accurately model the $\CP^1$ field dynamics. In fact, 
recent numerical \cite{linsad} and analytic \cite{hasspe,str} work gives 
some grounds for optimism in the equivariant $n\geq 2$ case.

Staying 
within the geodesic approximation, there are many interesting open 
questions about the $L^2$ geometry of $\rat_n$ which require a good 
understanding of its boundary at infinity, so far lacking except in the 
case $n=1$. For example, is the volume and/or diameter finite? Is the 
spectrum of the Laplacian continuous or discrete (the answer having 
implications for quantum lump dynamics)? In this paper we have obtained a 
comprehensive understanding of the boundary at infinity of a (very) low 
dimensional totally geodesic submanifold of $\rat_n$ which suggests that 
constructing natural $n$-fold covers of $\rat_n$ may be a productive line 
of attack.

\section*{Acknowledgements} JAM acknowledges the University of 
Leeds and EPSRC for financial support, and Zen Harper for useful 
conversations.

\end{document}